\documentclass[a4paper,11pt]{article}

\usepackage[dvips]{graphicx,psfrag}
\usepackage{amssymb}

\makeatletter

 \@addtoreset{equation}{section}
\makeatother
\newcommand{\beq}{\begin{eqnarray}}
\newcommand{\eeq}{\end{eqnarray}}
\def\tr{\mathop{\mathrm{tr}}\nolimits}

\begin{document}

\begin{titlepage}

\begin{flushright}
OU-HET 439\\
{\tt hep-th/0304194}\\
April 2003
\end{flushright}
\bigskip

\begin{center}
{\LARGE\bf
Three Dimensional Nonlinear Sigma Models
in the Wilsonian Renormalization Method
}
\vspace{1cm}

\setcounter{footnote}{0}
\bigskip

\bigskip
{\renewcommand{\thefootnote}{\fnsymbol{footnote}}
{\large\bf Kiyoshi Higashijima\footnote{
     E-mail: {\tt higashij@phys.sci.osaka-u.ac.jp}} and
 Etsuko Itou\footnote{
     E-mail: {\tt itou@het.phys.sci.osaka-u.ac.jp}}
}}

\vspace{4mm}

{\sl
Department of Physics,
Graduate School of Science, Osaka University,\\ 
Toyonaka, Osaka 560-0043, Japan \\
}

\end{center}
\bigskip

\begin{abstract}
The three dimensional nonlinear sigma model is nonrenormalizable within the perturbative method. Using the $\beta$ function in the nonperturbative Wilsonian renormalization group method, we argue that some ${\cal N}=2$ supersymmetric nonlinear $\sigma$ models are renormalizable in three dimensions.
When the target space is an Einstein-K\"{a}hler manifold with positive scalar curvature, such as C$P^N$ or $Q^N$, there are nontrivial ultraviolet (UV) fixed points, which can be used to define the nontrivial renormalized theory. 
If the target space has a negative scalar curvature, however, the theory has only an infrared Gaussian fixed point, and the meaningful continuum theory cannot be defined. We also construct a model that interpolates between the C$P^N$ and $Q^N$ models with two coupling constants. This model has two non-trivial UV fixed points that can be used to define a nontrivial renormalized theory. 
Finally, we construct a class of conformal field theories with ${\bf SU}(N)$ symmetry, defined at the fixed point of the nonperturbative $\beta$ function. These conformal field theories have a free parameter corresponding to the anomalous dimension of the scalar fields. If we choose a specific value of this parameter, we recover the conformal field theory defined at the UV fixed point of the C$P^N$ model, and the symmetry is enhanced to ${\bf SU}(N+1)$.

\end{abstract}

\end{titlepage}

\pagestyle{plain}

\section{Introduction}
Nonlinear sigma models (NL$\sigma$Ms) are renormalizable in two dimensions. They are nonrenormalizable, however, in three dimensions within the perturbation theory. Therefore, we have to use nonperturbative methods to study the renormalizability of NL$\sigma$Ms. One of such nonperturbative methods is the large-$N$ expansion, which has been applied to some examples of NL$\sigma$Ms \cite{Arefeva}. For example, the ${\bf U}(N)$ invariant NL$\sigma$M, in which field variables take values in the complex projective space $CP^{N-1}$, is renormalizable at leading and next-to-leading orders in the large-$N$ expansion \cite{Inami}. In fact, the next-to-leading order contribution to the $\beta$ function in the C$P^N$ model with ${\cal N}=2$ supersymmetry vanishes in the $1/N$ expansion \cite{Inami}.

The Wilsonian renormalization group (WRG) offers another powerful tool suitable for nonperturbative studies. The $2$- and $3$-dimensional NL$\sigma$Ms represent theories of superstrings and supermembranes, respectively, and the target spaces of these theories correspond to real space-time. Since the consistency of these models requires conformal invariance, it is important to investigate the fixed point of the NL$\sigma$M with ${\cal N}=2$ supersymmetry. In a previous paper, we used the WRG method to show that the fixed points of $2$-dimensional NL$\sigma$Ms provide novel conformal field theories whose target spaces are non-Ricci flat \cite{SU2dim}. These target spaces are interpreted as consistent backgrounds of superstrings in the presence of a dilaton \cite{KKL}.

In the WRG approach, the renormalizability of NL$\sigma$Ms is equivalent to the existence of a nontrivial continuum limit, $\Lambda \rightarrow \infty$. When the ultraviolet (UV) cutoff $\Lambda$ tends to infinity, we have to fine tune the coupling constant to the critical value at the UV fixed point, so as to keep the observable quantities finite. Therefore, it is important to show the existence of the UV fixed point without using the perturbation theory. 

The WRG equation describes the variation of the general effective action when the cutoff scale is changed. The most general effective action depends on infinitely many coupling constants, and in practice, we have to introduce some kind of symmetry and truncation to solve the WRG equation. The simplest truncation is the local potential approximation, in which only the potential term without a derivative is retained. In the ${\cal N}=2$ supersymmetric NL$\sigma$Ms that we consider here, the field variables take values in complex curved spaces called K\"{a}hler manifolds,  whose metrics are specified completely by the K\"{a}hler potentials. Since the higher derivative terms are irrelevant in the infrared (IR) region, we assume that we can safely drop such terms in an regions. Furthermore, we assume that the ${\cal N}=2$ supersymmetry is maintained in the renormalization procedure.  Then, non-derivative terms such as $\det{g}$, disappear through a remarkable cancellation between bosons and fermions, and our action is completely fixed by the K\"{a}hler potential. Thus, our truncation method in which higher derivative terms are ignored is the first nontrivial approximation for ${\cal N}=2$ supersymmetric nonlinear $\sigma$ models. With these assumptions, we previously derived the WRG equation for the $3$-dimensional NL$\sigma$M \cite{HI}, and we found that the $\beta$ function for the target metric is 
\beq
\beta (g_{i \bar{j}})=\frac{1}{2 \pi^2} R_{i \bar{j}}+\gamma [\varphi^k g_{i \bar{j},k} +\varphi^{* \bar{k}}g_{i \bar{j}, \bar{k}}+2 g_{i \bar{j}}  ]+\frac{1}{2}[\varphi^k g_{i \bar{j},k} +\varphi^{* \bar{k}}g_{i \bar{j}, \bar{k}}].
\eeq
In this paper, we investigate the $3$-dimensional ${\cal N}=2$ supersymmetric NL$\sigma$M using this equation.

First, we study the renormalization group flows for some simple models.
One of them is the theory whose target space is the Einstein-K\"{a}hler manifold, including the C$P^N$ and $Q^N$ models.
We show that if the target manifold has a positive scalar curvature, the theory has a nontrivial UV fixed point together with an IR fixed point at the origin. It is possible to take the continuum limit of this theory by using the UV fixed point, so that these NL$\sigma$Ms are renormalizable in three dimensions, at least within our truncation method. We also study a model whose target manifold is not the Einstein-K\"{a}hler manifold. This model has two coupling constants and connects the C$P^N$ and $Q^N$ models through the renormalization group flow. The theory spaces are divided into four phases.

Next, we construct the conformal field theories at the fixed point of the nonperturbative $\beta$ function in more general theory spaces. To simplify, we assume ${\bf SU}(N)$ symmetry for the theory and obtain a class of ${\bf SU}(N)$ invariant conformal field theories with one free parameter. If we choose a specific value of this free parameter, we obtain a conformal field theory defined at the UV fixed point of the C$P^N$ model and the symmetry is enhanced to ${\bf SU}(N+1)$ in this case.
The conformal field theory reduces to a free field theory if the free parameter is set to zero. Therefore, the free parameter describes a marginal deformation from the IR to the UV fixed point of the C$P^N$ model in the theory spaces.

This paper is organized as follows:
In \S \ref{review}, we give a short review of the derivation of the WRG equation for NL$\sigma$Ms.
In \S \ref{E-K}, we argue the renormalizability of ${\cal N}=2$ supersymmetric nonlinear $\sigma$ models whose target spaces are the Einstein-K\"{a}hler manifold with positive scalar curvature. 
In \S \ref{RGflow}, we study the renormalization group flow for some specific models.
We construct a class of conformal field theories with ${\bf SU}(N)$ symmetry in \S \ref{solution}.
Such a class has one free parameter, corresponding to the anomalous dimension of the scalar field.
In \S \ref{corres-CPN}, we show that the conformal theory is equivalent to the UV fixed point of the C$P^N$ model if we take a specific value of the parameter.


\section{Wilsonian Renormalization Group (WRG) equation}\label{review}

In general, the WRG eq. describes the variation of the effective action $S$ when the cutoff energy scale $\Lambda$ is changed to $\Lambda (\delta t)=\Lambda e^{-\delta t}$ in $D$-dimensional field theory \cite{Wilson Kogut, Wegner and Houghton, Morris, Aoki}:
\beq  
\frac{d}{dt}S[\Omega; t]&=&\frac{1}{2\delta t} \int_{p'} \tr \ln \left(\frac{\delta^2 S}{\delta \Omega^i \delta \Omega^j}\right)\nonumber\\
&&-\frac{1}{2 \delta t}\int_{p'} \int_{q'} \frac{\delta S}{\delta \Omega^i (p')} \left(\frac{\delta^2 S}{\delta \Omega^i (p')\delta \Omega^j (q')} \right)^{-1} \frac{\delta S}{\delta \Omega^j (q')} \nonumber\\
&&+ \left[D-\sum_{\Omega_i} \int_p \hat{\Omega}_i (p) \left(d_{\Omega_i}+\gamma_{\Omega_i}+\hat{p}^{\mu} \frac{\partial}{\partial \hat{p}^{\mu}} \right) \frac{\delta}{\delta \hat{\Omega}_i (p)} \right] \hat{S}.\nonumber\\ \label{WRG-1} 
\eeq
Here, $d_{\Omega}$ and $\gamma_{\Omega}$ denote the canonical and anomalous dimensions of the field $\Omega$, respectively.
The carat indicates dimensionless quantities.
The first and second terms in Eq.~(\ref{WRG-1}) correspond to the one-loop and dumbbell diagrams, respectively.
The remaining terms come from the rescaling of fields.
We always normalize the coefficient of the kinetic term to unity.

This WRG equation consists of an infinite set of differential equations for the various coupling constants within the most general action $S$.
In practice, we usually expand the action in powers of derivatives and retain the first few terms.
We often introduce some symmetry to further reduce the number of independent coupling constants.

We consider the simplest ${\cal N}=2$ supersymmetric theory, which is given by the K\"{a}hler potential term.
In this case, the action is written
\beq
S&=&\int d^2 \theta d^2 \bar{\theta} d^3 x K[\Phi, \Phi^\dag]\nonumber\\
&=&\int d^3 x \Bigg[g_{n \bar{m}}\left(\partial^{\mu} \varphi^n \partial_{\mu} \varphi^{* \bar{m}} +\frac{i}{2} \bar{\psi}^{\bar{m}} \sigma^{\mu}(D_{\mu} \psi)^n +\frac{i}{2} \psi^{n} \bar{\sigma}^{\mu}(D_{\mu} \bar{\psi})^{\bar{m}} +\bar{F}^{\bar{m}} F^{n}\right) \nonumber\\
&&-\frac{1}{2} K_{,nm \bar{l}} \bar{F}^{\bar{l}} \psi^n \psi^m -\frac{1}{2} K_{,n \bar{m} \bar{l}} F^{n} \bar{\psi}^{\bar{m}} \bar{\psi}^{\bar{l}}+\frac{1}{4} K_{,nm \bar{k} \bar{l}} (\bar{\psi}^{\bar{k}} \bar{\psi}^{\bar{l}})(\psi^n \psi^m)\Bigg],\label{action}
\eeq
where $\Phi^n$ represents chiral superfields, whose component fields are complex scalars $\varphi^n (x)$, Dirac fermions $\psi^n (x)$ and complex auxiliary fields $F^n (x)$.
In the usual perturbative method, this model is not renormalizable.
We examine it using the WRG equation, which constitutes one of nonperturbative methods.

When we substitute this action into Eq.(\ref{WRG-1}), the one-loop correction term cannot be written in covariant form \cite{HI}.
We use the K\"{a}hler normal coordinates (KNC) to obtain a covariant expression for the loop correction term. In order to integrate over the rapidly fluctuating fields, we have to expand the action around the slowly varying fields, which are treated as background fields. We transform the fluctuating fields to the KNC by holomorphic transformations \cite{HN}. Then the action for the fluctuating fields is expressed in terms of covariant quantities under the general coordinate transformations of the background fields. After integration over the fluctuating fields, the resulting effective action for the background fields is covariant under the general coordinate transformations. 
The Jacobian for the path integral measure cancels between bosons and fermions because of the supersymmetry \cite{HN2}.
In nonlinear sigma models defined on a coset manifold $G/H$, a part of the global symmetry is realized nonlinearly. If the general coordinate transformation of the target manifold allows an infinitesimal transformation that leaves the metric invariant, the generator of the transformation is called a Killing vector and defines the symmetry of the corresponding nonlinear sigma model. In order to maintain the global symmetries that are realized nonlinearly, it is desirable to respect covariance under the general coordinate transformation.

Finally, we obtain the WRG equation for the scalar part as
\beq
&&\frac{d}{dt}\int d^3 x g_{i \bar{j}} (\partial_\mu \varphi)^i (\partial^\mu \varphi^*)^{\bar{j}}\nonumber\\
&&=\int d^3 x \Big[-\frac{1}{2 \pi^2} R_{i \bar{j}} \nonumber\\
&&-\gamma \Big(\varphi^k g_{i \bar{j},k}+ \varphi^{*\bar{k}}g_{i \bar{j},\bar{k}} +2g_{i \bar{j}} \Big) -\frac{1}{2} \Big( \varphi^k g_{i \bar{j},k}+ \varphi^{*\bar{k}}g_{i \bar{j},\bar{k}} \Big) \Big] (\partial_\mu \varphi)^i  (\partial^\mu \varphi^*)^{\bar{j}},\label{WRG3dim} \nonumber\\
\eeq
where it has been assumed that the scalar fields $\varphi^n (x)$ are made independent of $t$ through a suitable rescaling, which introduces the anomalous dimension $\gamma$.
From this WRG equation, the $\beta$ function of the K\"{a}hler metric is
\beq
\frac{d}{dt}g_{i \bar{j}}&=&-\frac{1}{2 \pi^2}R_{i \bar{j}} -\gamma \Big[\varphi^k g_{i \bar{j},k} +\varphi^{* \bar{k}}g_{i \bar{j},\bar{k}}+2g_{i \bar{j}} \Big] -\frac{1}{2} \Big[\varphi^k g_{i \bar{j},k} +\varphi^{* \bar{k}}g_{i \bar{j},\bar{k}} \Big]  \nonumber\\
&\equiv&-\beta (g_{i \bar{j}}).\label{beta}
\eeq

\section{Einstein-K\"{a}hler manifolds}\label{E-K}
Let us consider the theories whose target spaces are Einstein-K\"{a}hler manifolds.
The Einstein-K\"{a}hler manifolds satisfy the condition
\beq
R_{i \bar{j}}=\frac{h}{a^2}g_{i \bar{j}},\label{EKcond}
\eeq
where $a$ is the radius of the manifold, which is related to the coupling constant $\lambda$ by
\beq
\lambda =\frac{1}{a}.
\eeq
There is a special class of Einstein-K\"{a}hler manifolds called the Hermitian symmetric space that consists of a symmetric coset space ($G/H$), namely, if the coset space is invariant under a parity operation. If the manifold is the Hermitian symmetric space, the positive constant $h$ in Eq.~(\ref{EKcond}) is the eigenvalue of the quadratic Casimir operator in  the adjoint representation of the global symmetry $G$, as shown in Table $1$ \cite{HKN}.

	
	\begin{table}[h]
	
	\begin{center}
	\begin{tabular}{|c|c|c|}
	\hline
	$G/H$                              & Dimensions (complex)  &$h$\\
	\hline \hline
	${\bf SU}(N)/[{\bf SU}(N-1) \otimes {\bf U}(1)]=$C$P^{N-1}$   & $N-1$        & $N$\\
	${\bf SU}(N)/[{\bf SU}(N-M) \otimes {\bf U}(M)]$   & $M(N-M)$                & $N$\\
	${\bf SO}(N)/[{\bf SO}(N-2) \otimes {\bf U}(1)]=Q^{N-2}$   & $N-2$           & $N-2$\\
	${\bf Sp}(N)/{\bf U}(N)$                     & $\frac{1}{2}N(N+1)$   & $N+1$\\
	${\bf SO}(2N)/{\bf U}(N)$                    & $\frac{1}{2}N(N+1)$   & $N-1$\\
	$E_{6}/[{\bf SO}(10) \otimes {\bf U}(1)]$    &$16$                     &$12$\\
	$E_{7}/[E_6 \otimes {\bf U}(1)]$        &$27$                     &$18$\\
	\hline
	\end{tabular}
	\caption{The values of $h$ for Hermitian symmetric spaces}
	\end{center}
	\end{table}
	

If the manifolds have radius $a=\frac{1}{\lambda}$, the scalar part of the SNL$\sigma$M Lagrangian can be represented in the following form:
\beq
{\cal L}_{scalar}&=&g_{i \bar{j}}(\varphi,\varphi^*) \partial_{\mu} \varphi^i \partial^{\mu} \varphi^{* \bar{j}} \nonumber\\
&\stackrel{\varphi,\varphi^* \approx 0}{\longrightarrow}& \frac{1}{\lambda^2} \delta_{i \bar{j}} \partial_{\mu} \varphi^i \partial^{\mu} \varphi^{* \bar{j}}.\label{lag-1}
\eeq
To normalize the kinetic term, we rescale the scalar fields as follows:
\beq
\varphi \rightarrow \tilde{\varphi}=\frac{1}{\lambda} \varphi.\label{rescale-1}
\eeq
Then, the Lagrangian (\ref{lag-1}) has the normalized kinetic term
\beq
{\cal L}_{scalar}=\tilde{g}_{i \bar{j}}(\lambda \tilde{\varphi},\lambda \tilde{\varphi}^*) \partial_{\mu} \tilde{\varphi}^i \partial^{\mu} \tilde{\varphi}^{* \bar{j}},
\eeq
with
\beq
\tilde{g}_{i \bar{j}}|_{\tilde{\varphi},\tilde{\varphi}^*=0}=\delta_{i \bar{j}}.
\eeq
Rescaling the WRG Eq.(\ref{WRG3dim}) and comparing the coefficients of the $\partial_{\mu} \tilde{\varphi}^i \partial^{\mu} \tilde{\varphi}^{* \bar{j}}$ terms, we have
\beq
\frac{\partial}{\partial t} \tilde{g}_{i \bar{j}} (\lambda \tilde{\varphi},\lambda \tilde{\varphi}^*)&=&-\frac{1}{2\pi^2} \tilde{R}_{i \bar{j}} \nonumber\\
&&-\gamma [\tilde{\varphi}^k \tilde{g}_{i \bar{j},k }+\tilde{\varphi}^{* \bar{k}} \tilde{g}_{i \bar{j},\bar{k}}+2 \tilde{g}_{i \bar{j}}]-\frac{1}{2}[\tilde{\varphi}^k \tilde{g}_{i \bar{j},k }+\tilde{\varphi}^{* \bar{k}} \tilde{g}_{i \bar{j},\bar{k}}],\nonumber
\eeq
where $\tilde{R}_{i \bar{j}}$ is the rescaled Ricci tensor, which can be written
\beq
\tilde{R}_{i \bar{j}}=h\lambda^2 \tilde{g}_{i \bar{j}}
\eeq
using the Einstein K\"{a}hler condition (\ref{EKcond}).
Because only $\lambda$ depends on $t$, this differential equation can be rewritten as
\beq
\frac{\dot{\lambda}}{\lambda} \tilde{\varphi}^k \tilde{g}_{i \bar{j},k } +\frac{\dot{\lambda}}{\lambda} \tilde{\varphi}^{* \bar{k}} \tilde{g}_{i \bar{j},\bar{k}}&=&-\left(\frac{h \lambda^2}{2\pi^2}+2\gamma  \right) \tilde{g}_{i \bar{j}}-(\gamma+\frac{1}{2})[\tilde{\varphi}^k \tilde{g}_{i \bar{j},k }+\tilde{\varphi}^{* \bar{k}} \tilde{g}_{i \bar{j},\bar{k}}].\nonumber\\. 
\eeq
The left-hand side vanishes for $\varphi,\varphi^* \approx 0$, so that the coefficient of $\tilde{g}_{i \bar{j}}$ on the right-hand side must vanish. Thus, we obtain the anomalous dimension of scalar fields (or chiral superfields) as
\beq
\gamma=- \frac{h \lambda^2}{4\pi^2}.\label{gamma}
\eeq
Then, comparing the coefficients of the $\tilde{\varphi}^k \tilde{g}_{i \bar{j},k }$ (or $\tilde{\varphi}^{* \bar{k}} \tilde{g}_{i \bar{j},\bar{k}}$) terms, we also obtain the $\beta$ function of $\lambda$:
\beq
\beta(\lambda)&\equiv&-\frac{d \lambda}{dt}=-\frac{h}{4\pi^2}\lambda^3+\frac{1}{2} \lambda .\label{beta-lambda}
\eeq
We have an IR fixed point at 
\beq
\lambda=0,
\eeq
and we also have a UV fixed point at 
\beq
\lambda^2=\frac{2 \pi^2}{h}\equiv \lambda_c^2
\eeq
for positive $h$. {\it Therefore, if the constant $h$ is positive, it is possible to take the continuum limit by choosing the cutoff dependence of the bare coupling constant as
\beq
\lambda(\Lambda) \stackrel {\Lambda \rightarrow \infty}{\longrightarrow} \lambda_c-\frac{M}{\Lambda},\label{continuum}
\eeq
where $M$ is a finite mass scale.}
With this fine tuning, ${\cal N}=2$ supersymmetric nonlinear $\sigma$ models are renormalizable, at least in our approximation, if the target spaces are Einstein-K\"{a}hler manifolds with positive curvature.

When the constant $h$ is positive, the target manifold is a compact Einstein-K\"{a}hler manifold \cite{Page and Pope}.
In this case, the anomalous dimensions at the fixed points are given by
\beq
\gamma_{IR}&=&0 \mbox{ at the IR fixed point (Gaussian fixed point),}\\
\gamma_{UV}&=&-\frac{1}{2} \mbox{at the UV fixed point.}
\eeq
At the UV fixed point, the scaling dimension of the scalar fields ($x_{\varphi}$) is equal to the canonical plus anomalous dimensions:
\beq
x_{\varphi}&\equiv& d_{\varphi} + \gamma_{\varphi}=0.
\eeq
Thus the scalar fields and the chiral superfields are dimensionless in the UV conformal theory, as in the case of two dimensional field theories.
Above the fixed point, the scalar fields have non-vanishing mass, and the symmetry is restored \cite{HKNT,HIT}.

\section{Renormalization group flows}\label{RGflow}
In this section, we study the renormalization group flows for three examples, C$P^N$, $Q^N$ and a new model.

\subsection{C$P^{N}$ and $Q^N$ models}
\begin{enumerate}
	\item C$P^N$ model: ${\bf SU}(N+1)/[{\bf SU}(N) \otimes {\bf U}(1)]$\\
	Consider the following ${\bf SU}(N+1)$ symmetric K\"{a}hler potential with $(N+1)$-dimensional homogeneous coordinates:
	\beq
	K[\Phi, \Phi^\dag]&=& \frac{1}{\lambda^2} \ln ( |\Phi^1|^2 + \cdots +|\Phi^N|^2 +|\Phi^{N+1}|^2).
	\eeq
	The complex projective space, C$P^N$, is defined by identifying two points related by
	\beq
	\Phi^i &\sim& a \Phi^i, \hspace{1cm} (i=1, \cdots, N+1)
	\eeq
	where $a$ is a complex chiral superfield, so that this is a complexified gauge symmetry. 	We obtain the K\"{a}hler potential for the C$P^N$ model by choosing the gauge $\Phi^{N+1}=1$ as
	\beq
	K[\Phi, \Phi^\dag]=\frac{1}{\lambda^2} \ln (1+ \vec{\Phi}\vec{\Phi}^{\dag}),
	\eeq
	where $\vec{\Phi}$ denotes a set of chiral superfields: $\vec{\Phi}=(\Phi^1, \cdots ,\Phi^{N})$.
	Hereafter, we rescale the scalar fields as
	\beq
	\varphi \rightarrow \tilde{\varphi} =\frac{1}{\lambda} \varphi\label{rescale},
	\eeq
	to normalize the kinetic term,
	and simply write the rescaled scalar fields $\tilde{\varphi}$ as $\varphi$.  
	
	From the above K\"{a}hler potential, we obtain the following K\"{a}hler metric and Ricci tensor:
	\beq
	g_{i \bar{j}}&\equiv& \partial_i \partial_{\bar{j}}K= \Big( \frac{\delta_{i \bar{j}}}{1+\lambda^2 \vec{\varphi} \vec{\varphi}^*}- \frac{\lambda^2 \varphi^i \varphi^{* \bar{j}}}{(1+\lambda^2 \vec{\varphi} \vec{\varphi}^*)^2} \Big),\\
	R_{i \bar{j}}&\equiv&-\partial_{\bar{j}} \partial_i (\ln \det g_{k \bar{l}})=(N+1) \lambda^2 g_{i \bar{j}}.\label{CPN-Ricci}
	\eeq
	Equation (\ref{CPN-Ricci}) shows that this target manifold is an Einstein-K\"{a}hler manifold with $h=N+1$.
	Hence, Eqs. (\ref{gamma}) and (\ref{beta-lambda}) immediately give us
	\beq
	\gamma&=&-\frac{(N+1)\lambda^2}{4 \pi^2},\\
	\beta (\lambda)&=& -\frac{(N+1)\lambda^3}{4 \pi^2}+\frac{1}{2}\lambda.
	\eeq
	This $\beta$ function is consistent with the large $N$ analysis \cite{Inami}.
	
	\item $Q^N$ model: ${\bf SO}(N+2)/[{\bf SO}(N) \otimes {\bf SO}(2)]$\\
	Another example of an Einstein-K\"{a}hler manifold is the coset manifold ${\bf SO}(N+2)/[{\bf SO}(N) \otimes {\bf SO}(2)]$ called $Q^N$.	We consider the K\"{a}hler potential with homogeneous $(N+2)$-dimensional coordinates
	\beq
	K[\Phi,\Phi^\dag]=\frac{1}{\lambda^2} \ln ( |\Phi^1|^2 + \cdots +|\Phi^N|^2 +|\Phi^{N+1}|^2+|\Phi^{N+2}|^2).
	\eeq
	Now, we impose the following identification and $O(N)$ symmetric conditions on the K\"{a}hler potential:
	\beq
	\Phi^i &\sim& a \Phi^i, \hspace{1cm} (i=1, \cdots, N+2),\\
	(\Phi^1)^2 + &\cdots& +(\Phi^N)^2 +(\Phi^{N+1})^2 +(\Phi^{N+2})^2=0,
	\eeq
	By these condition, the dimensions of the target space becomes $N$, and the K\"{a}hler potential for $Q^N$ can be rewritten \footnote{We choose the same gauge as in the next subsection.}
	\beq
	K[\Phi,\Phi^\dag]=\frac{1}{\lambda^2} \ln \Big( 1+\vec{\Phi} \vec{\Phi}^\dag +\frac{1}{4} \vec{\Phi}^2 \vec{\Phi}^{\dag 2} \Big),
	\eeq
	where $\vec{\Phi}=(\Phi^1, \cdots ,\Phi^N)$.
	Hereafter, we use the rescaled fields (\ref{rescale}).
	From the above K\"{a}hler potential, the K\"{a}hler metric and Ricci tensor are given by
	\beq
	g_{i \bar{j}}&=&\frac{\delta_{i \bar{j}}}{1+\lambda^2 \vec{\varphi} \vec{\varphi}^* +\frac{1}{4} \lambda^4 \vec{\varphi}^2 \vec{\varphi}^{* 2}}\nonumber\\
	&&+\frac{\lambda^2 \varphi^i \varphi^{* \bar{j}} \left(1+\lambda^2 \vec{\varphi} \vec{\varphi}^* \right)-\lambda^2 \left(\varphi_i^* \varphi_{\bar{j}} + \frac{1}{2}\lambda^2 \vec{\varphi}^2 \varphi_i^* \varphi^{* \bar{j}} + \frac{1}{2}\lambda^2 \vec{\varphi}^{* 2} \varphi^i \varphi_{\bar{j}} \right) }{ \left(1+\lambda^2 \vec{\varphi} \vec{\varphi}^* +\frac{1}{4}\lambda^4 \vec{\varphi}^2 \vec{\varphi}^{* 2}  \right)^2 }  , \label{Q-metric}\nonumber\\
	\\
	R_{i \bar{j}}&=&N \lambda^2 g_{i \bar{j}} \label{Q-Ricci}.
	\eeq
	Equation (\ref{Q-Ricci}) shows that this manifold is also an Einstein-K\"{a}hler manifold with $h=N$.
	Employing the same argument as in the case of the C$P^N$ model, we obtain the anomalous dimension and $\beta$ function for the coupling constant:
	\beq
	\gamma&=&-\frac{N \lambda^2}{4 \pi^2},\\
	\beta (\lambda)&=& -\frac{N \lambda^3}{4 \pi^2}+\frac{1}{2}\lambda.
	\eeq
	
	The next example shows that there are renormalization group flows that connect the C$P^N$ and $Q^N$ models.
	
	\end{enumerate}
	
	\subsection{A new model}
	
	\begin{enumerate}
	\item Construction
	
	Again, we consider the K\"{a}hler potential with homogeneous $(N+2)$-dimensional coordinates:
	\beq
	K[\Phi,\Phi^\dag]=\frac{1}{\lambda^2} \ln ( |\Phi^1|^2 + \cdots +|\Phi^N|^2 +|\Phi^{N+1}|^2+|\Phi^{N+2}|^2).\label{homo-kahler}
	\eeq
	As in the C$P^N$ and $Q^N$ models, we identify two points related by
	\beq
	\Phi^i &\sim& a \Phi^i. \hspace{1cm} (i=1, \cdots, N+2) \label{identify}
	\eeq
	Now, we deform the ${\bf O}(N)$ symmetric condition to
	\beq
	b[(\Phi^1)^2 + &\cdots& +(\Phi^N)^2] +(\Phi^{N+1})^2 +(\Phi^{N+2})^2=0,\label{deform-condition}
    \eeq
	where $b$ is an arbitrary complex parameter.

	\begin{enumerate}
		\item The case $b=0$: \\
		In this case, the deformed condition (\ref{deform-condition}) can be rewritten as
		\beq
		(\Phi^{N+1})^2 +(\Phi^{N+2})^2 =0.
		\eeq
		Here, we have fixed $\Phi^{N+1}$ and $\Phi^{N+2}$ by using the two conditions (\ref{identify}) and (\ref{deform-condition}) as follows:
		\beq
		\Phi^{N+1}&=&\frac{1}{{\sqrt{2}}},\\
		\Phi^{N+2}&=&\pm \frac{i}{\sqrt{2}}.
		\eeq
		Substituting these values into the K\"{a}hler potential (\ref{homo-kahler}), we obtain the K\"{a}hler potential of C$P^N$,
		\beq
		K[\Phi, \Phi^\dag]=\frac{1}{\lambda^2} \ln (1 + |\Phi^1|^2 + \cdots +|\Phi^N|^2).
		\eeq
		
		Thus, the target space is a double cover of C$P^N$ located at $\Phi^{N+2}= \pm \frac{i}{\sqrt{2}}$.
		This target manifold has isometry ${\bf SU}(N+1)$.

		\item The case $b \neq 0$: \\
		Here, using the two conditions, we can choose the gauge
		\beq
		\Phi^{N+1}+i \Phi^{N+2}=\sqrt{2},
		\eeq
		and
		\beq
		\Phi^{N+1} -i \Phi^{N+2}=\frac{-b}{\sqrt{2}} \Big( (\Phi^1)^2 +\cdots +(\Phi^{N})^2 \Big).
		\eeq
		Then, the K\"{a}hler potential can be rewritten
		\beq
		K[\Phi, \Phi^\dag]=\frac{1}{\lambda^2} \ln \Big( 1+|\Phi^1|^2 +\cdots +|\Phi^{N}|^2 +\frac{|b|^2}{4} | \sum^N_{i=1} (\Phi^i)^2 |^2  \Big).\label{new-kahler} \nonumber\\
		\eeq
		If we take $|b|=1$, this K\"{a}hler potential is identical to that of the $Q^N$ model.
		Thus for this special value of $b$, the target manifold has isometry ${\bf SO}(N+2)$.

		\item The case $b=\infty$: \\
		In this case, the ${\bf O}(N)$ symmetric condition reduces to
		\beq
		(\Phi^1)^2 + \cdots +(\Phi^N)^2=0.
		\eeq
		The remaining fields $\Phi^{N+1}$ and $\Phi^{N+2}$ can take arbitrary values.
		Using the identification condition, we can fix 
		\beq
		\Phi^{N-1} + i \Phi^{N}=\sqrt{2}.
		\eeq
		Then, we obtain the K\"{a}hler potential as follows:
		\beq
		&&K[\Phi,\Phi^\dag]\nonumber\\
		&&\hspace{-1cm}=\frac{1}{\lambda^2} \ln \Big( 1+|\Phi^1|^2 +\cdots +|\Phi^{N-2}|^2 +|\Phi^{N+1}|^2 +|\Phi^{N+2}|^2 + \frac{1}{4} |\sum^{N-2}_{i=1} (\Phi^i)^2 |^2 \Big).\label{b=infty}\nonumber\\
		\eeq

		\end{enumerate}
	
	\item Strong-weak duality\\
	There exists a very interesting duality for $N=2$.
	For $b=0$, the target space consists of a double cover of C$P^2$.
	For $|b|=1$, the target space is $Q^2$, which is isomorphic to C$P^1 \times$ C$P^1$.

	For $b=\infty$, if we choose 
	\beq
	\Phi^1 = \frac{1}{\sqrt{2}},\hspace{1cm}\Phi^2=\pm \frac{i}{\sqrt{2}},
	\eeq
	the K\"{a}hler potential (\ref{b=infty}) reads
	\beq
	K[\Phi,\Phi^\dag]=\frac{1}{\lambda^2} \ln (1+|\Phi^3|^2+|\Phi^4|^2),
	\eeq
	which is the K\"{a}hler potential of C$P^2$.
	Therefore, the target space is again a double cover of C$P^2$, and it coincides with that in the $b=0$ case.
	
	Let us replace the coordinates $\Phi^1 and \Phi^2$ with $\Phi^3 and \Phi^4$ in the constraint (\ref{deform-condition}). With this operation, the deformation parameter $b$ is replaced by $1/b$
	\beq
	b &\leftrightarrow& \frac{1}{b}.
	\eeq
Although the K\"{a}hler potential (\ref{new-kahler}) for $N=2$ is completely different for the deformation parameters $b$ and $1/b$, these two theories are equivalent.	
	Thus, this model for $N=2$ possesses strong-weak duality.
	The strong coupling region of the new model with (\ref{new-kahler}) corresponds to the weak coupling region of the dual model.
	At the self-dual point $|b|=1$, we have a model on $Q^2 \simeq $ C$P^1 \times $ C$P^1$. At $b=0 and \infty$, the target space of this theory is a double cover of C$P^2$.
	 
	\item Renormalization group flow
	
	Now, we study the renormalization group flow for general values of $b$.
	We use the K\"{a}hler potential 
	\beq
	K[\Phi,\Phi^\dag]=\frac{1}{\lambda^2} \ln \Big( 1+\vec{\Phi} \vec{\Phi}^\dag + g \vec{\Phi}^2 \vec{\Phi}^{\dag 2} \Big),
	\eeq
	where $g=\frac{|b|^2}{4}$ in Eq.~(\ref{new-kahler}).
	This K\"{a}hler potential yields the following K\"{a}hler metric and Ricci tensor:
\beq
g_{i \bar{j}}&=&\frac{\delta_{i \bar{j}}}{1+\lambda^2 \vec{\varphi} \vec{\varphi}^* +g(t)\lambda^4 \vec{\varphi}^2 \vec{\varphi}^{*2}}\nonumber\\
&&+\frac{4 g(t)\lambda^2 \varphi^i \varphi^{*\bar{j}}(1+\lambda^2 \vec{\varphi} \vec{\varphi}^*)-\lambda^2(\varphi_i^* \varphi_{\bar{j}} +2g(t)\lambda^2 \vec{\varphi}^2 \varphi_i^* \varphi^{*\bar{j}}+2g(t)\lambda^2 \vec{\varphi}^{*2}\varphi^i \varphi_{\bar{j}})}{(1+\lambda^2 \vec{\varphi} \vec{\varphi}^* +g(t) \lambda^4 \vec{\varphi}^2 \vec{\varphi}^{*2})^2},\nonumber\\
R_{i \bar{j}}&=&(N+1)\lambda^2  g_{i \bar{j}}-\Bigg[\frac{4g(t)\lambda^2 \delta_{i \bar{j}}}{1+4g(t)\lambda^2 \vec{\varphi} \vec{\varphi}^* +g(t)\lambda^4 \vec{\varphi}^2 \vec{\varphi}^{*2}} \nonumber\\
&&+\frac{16g^2 (t)\lambda^4 \varphi^i \varphi^{* \bar{j}} \vec{\varphi} \vec{\varphi}^{*} -16 g^2 (t) \lambda^2 \varphi_i^* \varphi_{\bar{j}} -8g^2 (t)\lambda^4 \left(\vec{\varphi}^2 \varphi_i^* \varphi^{*\bar{j}}+\vec{\varphi}^{*2}\varphi^i \varphi_{\bar{j}} \right)}{(1+4g(t)\lambda^2 \vec{\varphi} \vec{\varphi}^* +g(t)\lambda^4 \vec{\varphi}^2 \vec{\varphi}^{*2})^2} \Bigg].\label{new-Ricci}\nonumber\\
\eeq
Here, we have used the rescaled fields as before.
Note that Eq.(\ref{new-Ricci}) shows that this manifold is not an Einstein K\"{a}hler manifold unless $g$ takes the specific values $0$ or $\frac{1}{4}$.

	Substituting the above metric and Ricci tensor in Eq.(\ref{beta}), we obtain
	\beq
	\gamma&=&-\frac{\lambda^2}{4 \pi^2} [(N+1)-4g],\\
	\beta(\lambda)&=&-\frac{\lambda^3}{4 \pi^2}[(N+1) +8g(2g-1)]+\frac{\lambda}{2},\\
	\beta(g)&=&\frac{4 \lambda^2}{\pi^2} g^2 (4g-1).
	\eeq
	
\begin{figure}[h]
\begin{center}
\psfrag{g}{\Large$g$}
\psfrag{lambda}{\Large$\lambda$}
\includegraphics[width=12cm]{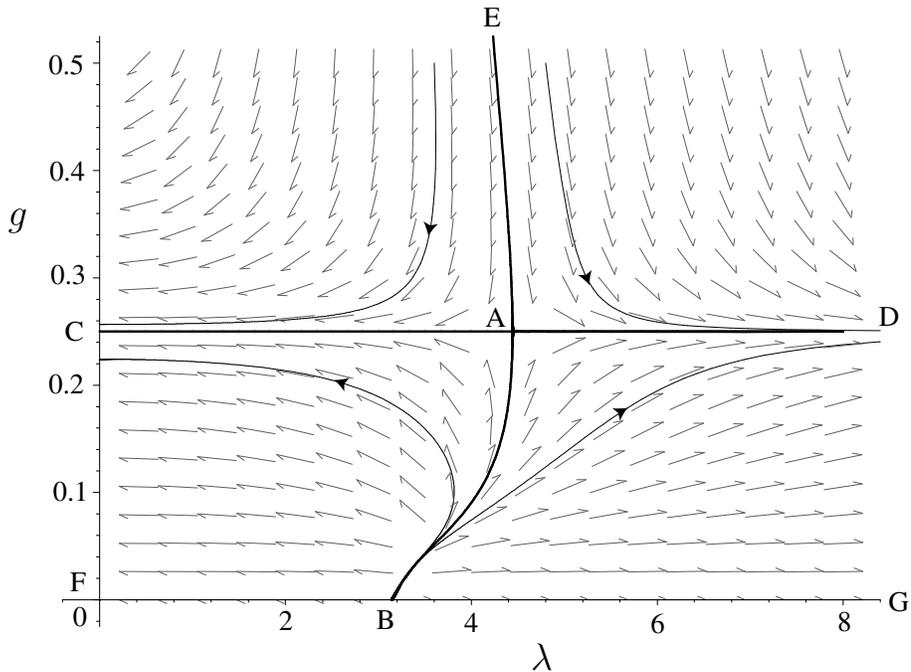}
\caption{Renormalization group flows (The arrows point toward the infrared region.)}
\label{fig:flow}
\end{center}
\end{figure}
	
	Figure \ref{fig:flow} displays the renormalization group flow.  If we use the perturbation theory, the flow diagram is reliable only in the vicinity of the origin. However, because we do not use the perturbation theory to derive the $\beta$ function, our flow diagram is reliable over the entire region. The nontrivial UV fixed points of the flow are the points labeled A and B. Any points on the $g$-axis ($\lambda=0$) are IR fixed points.  The curve BAE represents the critical line, along which the direction of the flow is tangent. The lines FBG ($g=0$) and CAD ($g=1/4$) both represent renormalized trajectories. The critical line intersects the renormalized trajectories at UV fixed points. We can define the continuum theories using these UV fixed points. In this sense, NL$\sigma$Ms are renormalizable in three dimensions, at least in the case of our truncated WRG equation.
	
	The theory has different symmetries on the right- and left-sides of the critical line.
	First, we consider continuous symmetry.
	The global symmetries on the renormalized trajectories, FBG ($g=0$) and CAD ($g=1/4$), are enhanced to $G=${\bf SU}$(N+1)$ and $G=${\bf SO}$(N+2)$, respectively.  In the other region of the flow diagram, the global symmetry is {\bf SO}$(N)\otimes ${\bf U}$(1)$. This global symmetry is realized manifestly on the right of the critical line BAE.  On the left of this critical line, however, the enhanced global symmetries are spontaneously broken, and there exist Nambu-Goldstone bosons, although the {\bf SO}$(N)\otimes ${\bf U}$(1)$ symmetry remains manifest. 
	Next, we consider the discrete transformation 
	\beq
	\psi(t,x_1,x_2) &\rightarrow& \psi' (t,x_1,x_2) =\gamma^2 \psi(t, x_1,-x_2),\nonumber\\
	x_2 &\rightarrow& x_2'=-x_2. \label{parity}
	\eeq
	The Lagrangian (\ref{action}) is invariant under this transformation.
	This transformation forbids fermion mass terms, so that it is broken spontaneously allows the massive phase.
	On the right of the critical line, the fermion is massive, and the discrete symmetry is spontaneously broken. On the left of the critical line, the discrete symmetry protects the fermion to be massless, and also, the supersymmetry keeps the bosons massless \cite{HIT}.
	
\end{enumerate}

\section{The SU(N) symmetric solution of the WRG equation}\label{solution}
In this section, we investigate the conformal field theories defined as the fixed points of the $\beta$ function
\beq
\beta&=&\frac{1}{2 \pi^2}R_{i \bar{j}} +\gamma \Big[\varphi^k g_{i \bar{j},k} +\varphi^{* \bar{k}}g_{i \bar{j},\bar{k}}+2g_{i \bar{j}} \Big] +\frac{1}{2} \Big[\varphi^k g_{i \bar{j},k} +\varphi^{* \bar{k}}g_{i \bar{j},\bar{k}} \Big]\nonumber\\
&=&0.\label{beta2} 
\eeq
To simplify, we assume the ${\bf S}{\bf U}(N)$ symmetric K\"{a}hler potential
\beq
K[\Phi,\Phi^\dag]&=&\sum_{n=1}^{\infty} g_n (\vec{\Phi} \cdot \vec{\Phi}^\dag)^n \equiv f(x),\label{potential}
\eeq
where the chiral superfields $\vec{\Phi}$ have $N$ components, $g_n$ plays the role of the coupling constant, and $x \equiv \vec{\Phi} \cdot \vec{\Phi}^\dag$.
Using the function $f(x)$, we derive the K\"{a}hler metric and Ricci tensor as follows:
\beq
g_{i \bar{j}}&\equiv&\partial_i \partial_{\bar{j}} K[\Phi,\Phi^\dag]=f' \delta_{i \bar{j}}+f'' \varphi_i^* \varphi_{\bar{j}},\label{metric}\\
R_{i \bar{j}}&\equiv&-\partial_i \partial_{\bar{j}} \tr \ln g_{i \bar{j}}\nonumber\\
&=&-\Big[(N-1)\frac{f''}{f'} +\frac{2f''+f''' x}{f'+f'' x}  \Big]\delta_{i \bar{j}} \nonumber\\
&&-\Big[(N-1)\bigg(\frac{f^{(3)}}{f''}-\frac{(f'')^2}{(f')^2} \bigg)+\frac{3f^{(3)}+f^{(4)} x}{f'+f'' x} -\frac{(2f''+f'''x)^2}{(f'+f''x)^2} \Big]\varphi^*_{i}\varphi_{\bar{j}},\nonumber\\
\eeq
where 
\beq
f'=\frac{df}{dx}.
\eeq
To normalize the kinetic term, we set 
\beq
f'|_{x \approx 0}=1
\Rightarrow g_1=1. \label{normalization}
\eeq
We substitute the above metric and Ricci tensor into the $\beta$ function (\ref{beta2}) and compare the coefficients of $\delta_{i \bar{j}}$ and $\varphi^i \varphi^{* \bar{j}}$, respectively. This yields
\beq
\frac{\partial}{\partial t}f'&=&\frac{1}{2\pi^2}\Big[(N-1)\frac{f''}{f'} +\frac{2f''+f''' x}{f'+f'' x}  \Big]- 2\gamma(f'+f''x)\label{f'}-f''x,\label{beta-f}\\
\frac{\partial}{\partial t}f''&=&\frac{1}{2\pi^2}\Big[(N-1)\bigg(\frac{f^{(3)}}{f''}-\frac{(f'')^2}{(f')^2} \bigg)+\frac{3f^{(3)}+f^{(4)} x}{f'+f'' x} -\frac{(2f''+f'''x)^2}{(f'+f''x)^2} \Big]\nonumber\\
&&-2\gamma(2f''+f'''x)-(f'''x+f'').\label{f''}
\eeq
Equation (\ref{f''}) is equivalent to the derivative of the first equation with respect to $x$. Therefore we use only the first equation.

To obtain a conformal field theory, we must solve the differential equation
\beq
\frac{\partial}{\partial t}f'=\frac{1}{2\pi^2}\Big[(N-1)\frac{f''}{f'} +\frac{2f''+f''' x}{f'+f'' x}  \Big]- 2\gamma(f'+f''x)\label{f'}-f''x =0\label{beta=0}.
\eeq
The function $f(x)$ is a polynomial of infinite degree, and it is difficult to solve it analytically. For this reason, we truncate the function $f(x)$ at order $O(x^4)$.
From the normalization (\ref{normalization}), the function $f(x)$ is 
\beq
f(x)=x+g_2 x^2 +g_3 x^3 +g_4 x^4.\label{order4}
\eeq
We substitute this into the WRG eq.~(\ref{beta-f}) and expand it about $x \approx 0$. Then Eq.~(\ref{beta=0}) can be written 
\beq
\frac{\partial}{\partial t}f'&=&\frac{1}{2 \pi^2} \Big[2(N+1)g_2 + \Big( 6(N+2)g_3 -4(N+3)g_2^2 \Big)x \nonumber\\
&&-\Big(18(N+7)g_2 g_3 -8(N+7)g_2^3 -12 (N+2)g_4 \Big)x^2  \Big]\nonumber\\
&&-2\gamma(1+4g_2 x +9 g_3 x^2 )-(2g_2 x +6 g_3 x^2 ) +O(x^3)\nonumber\\
&=&0.
\eeq
We choose the coupling constants and the anomalous dimension that satisfy this equation:
\beq
\gamma&=&\frac{N+1}{2 \pi^2} g_2,\\
g_3&=&\frac{2(3N+5)}{3(N+2)}g_2^2 +\frac{2\pi^2}{3(N+2)} g_2,\label{g_3}\\
g_4&=&3g_2 g_3 -\frac{2(N+7)}{3(N+3)}g_2^3 +\frac{\pi^2}{N+3}g_3\nonumber\\
&=&\frac{1}{3(N+2)(N+3)}\Big( (16N^2 +66N+62)g_2^3 + 2\pi^2 (6N+14)g_2^2+2\pi^4 g_2 \Big).\nonumber\\
\label{g_4}
\eeq
Note that all the coupling constants are written in terms of $g_2$ only.
Similarly, we can fix all the coupling constants $g_n$ using $g_2$ order by order.
The function $f(x)$ with such coupling constants describes the conformal field theory and has one free parameter, $g_2$.
In other words, if we fix the value of $g_2$, we obtain a conformal field theory.

\subsection{An explicit example of novel conformal field theories}\label{corres-CPN}
For general $g_2$, the power series (\ref{order4}) is a complicated function.
In this subsection, we consider a specific value of $g_2$, for which $f(x)$ takes an especially simple form.

We choose 
\beq
g_2&=&-\frac{1}{2} \cdot \frac{2\pi^2}{N+1} \equiv -\frac{1}{2}a,
\eeq
where
\beq
a\equiv \frac{2\pi^2}{N+1}.
\eeq
We can express all other coupling constants from this, using Eqs.(\ref{g_3}) and (\ref{g_4}). We have
\beq
g_3&=&\frac{1}{3} a^2,\nonumber\\
g_4&=&-\frac{1}{4}  a^3,\nonumber\\
\vdots.\nonumber
\eeq
These coupling constants show that the function $f(x)$ is 
\beq
f(x) =\frac{1}{a } \ln (1+ a x ),\label{CP-UV}
\eeq
and this is the K\"{a}hler potential of the C$P^N$ model.
In fact, the function (\ref{CP-UV}) satisfies the condition (\ref{beta=0}) exactly.

From this discussion, we find that one of the novel ${\bf SU}(N)$ symmetric conformal field theories is identical to the UV fixed point theory of the C$P^N$ model for a specific value of $g_2$.
In this case, the symmetry of this theory is enhanced to ${\bf SU}(N+1)$ because the C$P^N$ model has the isometry ${\bf SU}(N+1)$.

\section{Summary and Discussion}
The nonlinear $\sigma$ model in three dimensions is nonrenormalizable in perturbation theory.
We have studied ${\cal N}=2$ supersymmetric NL$\sigma$Ms using WRG equation, which is one of nonperturbative methods. 

First, we examined the sigma models whose target spaces are the Einstein-K\"{a}hler manifolds.
We have shown that the theories whose target spaces are compact Einstein-K\"{a}hler manifolds with positive scalar curvature have two fixed points.
One of them is the Gaussian IR fixed point and the other is the nontrivial UV fixed point. We can define the continuum limit at this UV fixed point by the fine-tuning of the bare coupling constant. In this sense, NL$\sigma$Ms on Einstein-K\"{a}hler manifolds with positive scalar curvature are renormalizable in three dimensions. At this point, the scaling dimension of all superfields is zero, as in the two dimensional theories. On the other hand, the theories whose target spaces are Einstein-K\"{a}hler manifolds with negative scalar curvature (for example $D^2$ with the Poincar\'{e} metric) have only an Gaussian IR fixed point, and cannot have a continuum limit.

Second, we studied a new model with two parameters, whose target space is not an Einstein-K\"{a}hler manifold. This theory has two nontrivial fixed points, corresponding to the UV fixed points of the C$P^N$ and $Q^N$ models.
In the theory spaces of this model, there are a critical surface and two renormalized trajectories, and the theory has four phases.
We have also shown that the model possesses strong-weak duality for $N=2$. In order to study the phase structure, we have to introduce the auxiliary fields \cite{HN3}. This is left for a future work.

Finally, we constructed a class of the ${\bf SU}(N)$ symmetric conformal field theory using the WRG equation. This class has one free parameter, $g_2$, corresponding to the anomalous dimension of the scalar fields. If we choose a specific value of this parameter, we recover the conformal field theory defined at the UV fixed point of the C$P^N$ model and the symmetry is enhanced to ${\bf SU}(N+1)$.

We argued that a certain class of ${\cal N}=2$ supersymmetric NL$\sigma$Ms are renormalizable in three dimensions. A similar argument may also be valid for ${\cal N}=1$ supersymmetric NL$\sigma$Ms and NL$\sigma$Ms without supersymmetry. Especially, NL$\sigma$Ms for Einstein manifolds with positive scalar curvature might be renormalizable, as long as we use Riemann normal coordinates.

\section*{Acknowledgements}
This work was supported in part by Grants-in-Aid for Scientific
Research (\#13640283 and \#13135215) and a Sasakawa Scientific Research Grant from The Japan Science Society. We would like to thank Muneto Nitta, Tetsuji Kimura and Makoto Tsuzuki for enlightening discussions.


\end{document}